\documentclass[amsmath,amssymb,11pt,superscriptaddress,reprint, preprintnumbers, notitlepage,aps,twocolumn,nofootinbib]{revtex4-1}
\pdfoutput=1
\usepackage[utf8]{inputenc} 
\usepackage[english]{babel}
\usepackage{amsmath}
\usepackage{graphicx}
\usepackage{dcolumn}
\usepackage{pbox}
\usepackage{amssymb}
\usepackage{epsfig}
\usepackage{slashed}
\usepackage{amssymb}
\usepackage{ mathrsfs }
\usepackage{color}
\usepackage{natbib}
\usepackage[font=small]{caption}
\usepackage[font=small]{subcaption}
\usepackage{url}
\usepackage{multirow}

\usepackage{lineno}
\usepackage{xcolor,pifont}
\usepackage{comment}

\definecolor{MyLightBlue}{rgb}{0.22,0.51,0.9}
\definecolor{BrickRed}{rgb}{0.8, 0.25, 0.33}
\RequirePackage{hyperref}
\hypersetup{colorlinks, citecolor=blue,linkcolor=red, urlcolor=MyLightBlue}

\makeatletter
\renewcommand\@makecaption[2]{%
  \par
  \vskip\abovecaptionskip
  \begingroup
  
   \small\rmfamily
    \begingroup
     \samepage
     \flushing
     \let\footnote\@footnotemark@gobble
     \@make@capt@title{#1}{#2}\par
    \endgroup
  \endgroup
  \vskip\belowcaptionskip
}
\makeatother

\DeclareUnicodeCharacter{2212}{-}
\begin{document}

\title{\bf Gravitational waves from metastable cosmic strings in Pati-Salam model\\ in light of new pulsar timing array data
}

\author{Waqas Ahmed}
\email[E-mail: ]{waqasmit@hbpu.edu.cn}
\affiliation{Hubei Polytechnic University, Huangshi 435003, China}

\author{Talal  Ahmed  Chowdhury}
\email[E-mail: ]{talal@du.ac.bd}
\affiliation{Department of Physics, University of Dhaka, P.O. Box 1000, Dhaka, Bangladesh}
\affiliation{Department of Physics and Astronomy, University of Kansas, Lawrence, Kansas 66045, USA}
\affiliation{The Abdus Salam International Centre for Theoretical Physics, Strada Costiera 11, I-34014, Trieste, Italy}

\author{Salah Nasri}
\email[E-mail: ]{snasri@uaeu.ac.ae; salah.nasri@cern.ch}
\affiliation{Department of Physics, UAE University, P.O. Box 17551, Al-Ain, United Arab Emirates}
\affiliation{The Abdus Salam International Centre for Theoretical Physics, Strada Costiera 11, I-34014, Trieste, Italy}

\author{Shaikh Saad}
\email[E-mail: ]{shaikh.saad@unibas.ch}
\affiliation{Department of Physics, University of Basel, Klingelbergstrasse\ 82, CH-4056 Basel, Switzerland}

\begin{abstract}
A series of pulsar timing arrays (PTAs) recently observed gravitational waves at the nanohertz frequencies. Motivated by this remarkable result, we present a novel class of  Pati-Salam models that give rise to a network of metastable cosmic strings, offering a plausible explanation for the observed PTA data. Besides, we introduce a hybrid inflationary scenario to eliminate magnetic monopoles that arise during the subsequent phase transitions from the Pati-Salam symmetry to the Standard Model gauge group. The resulting scalar spectral index is compatible with Planck data, and the tensor-to-scalar ratio is anticipated to be extremely small. Moreover, we incorporate a non-thermal leptogenesis to generate the required baryon asymmetry in our framework. Finally, the gravitational wave spectra generated by the metastable cosmic strings not only correspond to signals observed in recent PTAs, including NANOGrav, but are also within the exploration capacity of both present and future ground-based and space-based experiments.

\end{abstract}

\maketitle
\section{Introduction}
Gravitational waves (GWs) provide a unique window to probe fundamental physics. Recently, the International Pulsar Timing Array (IPTA) collaboration presented convincing evidence of such isotropic stochastic GW with frequency in the nanohertz range \cite{NANOGrav:2023gor, EPTA:2023fyk, Reardon:2023gzh, Xu:2023wog}. Although this stochastic GW background can be formed from the culmination of GWs produced from the in-spiraling and merging of the supermassive black holes in the universe~\cite{NANOGrav:2023hfp}, there is room for new physics (NP) explanation of such signal. Indeed, as pointed out in~\cite{NANOGrav:2023hvm,EPTA:2023xxk}, the GWs produced from metastable cosmic strings are compatible~\cite{Antusch:2023zjk, Buchmuller:2023aus, Fu:2023mdu, Lazarides:2023rqf, Ahmed:2023rky, Maji:2023fhv, afzal2023supersymmetric} with the recent results (for works on GWs, in light of previous PTA data, arising from cosmic strings, c.f., Refs.~\cite{Buchmuller:2019gfy,Dror:2019syi,King:2020hyd, Buchmuller:2020lbh, Buchmuller:2021dtt, Buchmuller:2021mbb, Dunsky:2021tih, Masoud:2021prr, Ahmed:2022rwy, Afzal:2022vjx, King:2021gmj,Lazarides:2022jgr,Fu:2022lrn,Saad:2022mzu,Maji:2023fba}). Cosmic strings can form in the early universe during the intermediate step(s) in the spontaneous symmetry breaking of a unifying gauge group to the Standard Model~\cite{Kibble:1976sj}. In this work, we investigate the possible production of metastable cosmic strings during the spontaneous symmetry breaking of the Pati-Salam gauge group~\cite{Pati:1973rp,Pati:1974yy}, $SU(4)_{C}\times SU(2)_{L}\times SU(2)_{R}$ to the SM. The Pati-Salam model introduced the attractive idea of the quark-lepton unification and also incorporated the left-right symmetry~\cite{Mohapatra:1974hk, Mohapatra:1974gc, Senjanovic:1975rk}. Besides, the model, automatically containing the right-handed neutrinos, can explain the smallness of the neutrino mass via the seesaw mechanism~\cite{Minkowski:1977sc, Gell-Mann:1979vob, Glashow:1979nm, Mohapatra:1979ia, Yanagida:1979as} and encompasses the leptogenesis mechanism~\cite{Fukugita:1986hr} to account for the matter-antimatter asymmetry of the universe. Additionally, the model naturally allows for neutron-antineutron oscillation~\cite{Mohapatra:1980qe}.

Particularly, we consider supersymmetric (SUSY) Pati-Salam model~\cite{Antoniadis:1988cm, King:1997ia, Jeannerot:2000sv, Melfo:2003xi} where we show that three viable symmetry breaking scenarios can lead to metastable cosmic string networks capable of explaining the PTA results. Among these three possibilities, we focus on a particular model where the Pati-Salam symmetry is first broken to the left-right symmetry, $SU(3)_{C}\times SU(2)_{L}\times SU(2)_{R}\times U(1)_{B-L}$. In the next phase transition, breaking of $SU(2)_R\to U(1)_R$ generates superheavy monopoles. In the subsequent breaking, i.e., when the last intermediate symmetry breaks down to that of the SM gauge group, the phase transition associated with  $U(1)_R\times U(1)_{B-L}\to U(1)_Y$ leads to cosmic string formation. Within this setup, we implement supersymmetric hybrid inflation at the last intermediate symmetry breaking scale, hence efficiently inflating away monopoles but not the cosmic strings. If these last two symmetry breaking scales almost overlap, metastable cosmic string networks are formed through the Schwinger nucleation. We find that the inflationary scenario incorporates the scalar spectral index consistent with Planck data, and the tensor-to-scalar ratio remains tiny. Interestingly, the new PTA data suggests that both the monopole and string formation scales must be close to $10^{15}$ GeV, which perfectly coincides with the seesaw scale, as well as provides the inflation scale and leads to successful non-thermal leptogenesis. Several gravitational wave observatories will fully test the stochastic gravitational wave background generated by the metastable string network.

The article is structured as follows. In Sec.~\ref{sec:2}, we discuss the recent PTA data and the production mechanism of metastable cosmic string networks and. In Sec.~\ref{sec:3}, we introduce a class of Pati-Salam models leading to the formation of metastable cosmic string networks, and in Sec.~\ref{sec:4} we fully construct one of the viable candidate models. In Sec.~\ref{sec:5}, we provide the details of the SUSY hybrid inflation and give a detailed account of non-thermal leptogenesis arising in our scenario. Finally we present the results in Sec.~\ref{CMB} and conclude in Sec.~\ref{sec:con}.

\section{Metastable Cosmic Strings \\and PTA data}\label{sec:2}

Cosmic strings are one-dimensional topological defects that arise when an Abelian symmetry is spontaneously broken. In this context, we use the Nambu-Goto string approximation, which assumes that the primary mode of radiation emission is in the form of GWs~\cite{Vachaspati:1984gt}.

The macroscopic properties of these cosmic strings are defined by their energy per unit length, denoted as $\mu_{cs}$, and referred to as the string tension. In our study, we explore models where the breaking of the Abelian symmetry is linked to the vacuum expectation value (vev) of the multiplets that also give rise to masses to the RHNs. Consequently, the tension of the cosmic strings are determined by the corresponding symmetry breaking scale,
\begin{align}
\mu_{cs}\sim 2\pi M^2,     
\end{align}
where order one coefficient is not shown explicitly (for details, see Ref.~\cite{Hill:1987qx}). 
In the above equation, $M$ refers to the symmetry breaking scale that creates the cosmic string network.   

On the contrary, when a simple group is broken down into a subgroup that includes an Abelian factor, it gives rise to the creation of monopoles~\cite{tHooft:1974kcl, Polyakov:1974ek}. However, to avoid the problem of overclosing the universe, inflationary processes must eliminate these monopoles. Subsequently, in a later stage, after the remaining Abelian symmetry is broken, cosmic strings emerge. When the scales of monopoles and cosmic strings are in close proximity, Schwinger nucleation occurs, leading to the creation of monopole-antimonopole pairs~\cite{Langacker:1980kd,Lazarides:1981fv,Vilenkin:1982hm} on the string, causing it to decay. In such a scenario, at high frequency regime, the string behaves like a stable one, whereas, at lower frequency regime, its behavior deviates form stable strings. When these metastable strings decay depends on the ratio of the monopole and string formation scales, 
\begin{align}
\alpha= \frac{m_{MP}^2}{\mu_{cs}}\sim \frac{8\pi}{g^2} \left( \frac{M^\prime}{M} \right)^2,   \label{eq:alpha} 
\end{align}
Here, $m_{MP}$ denotes the mass of the monopole,  $M^\prime$ represents the monopole creation scale, and $g$ stands for the relevant gauge coupling constant. When $\sqrt{\alpha}\gg 9$, the network exhibits behavior akin to that of a stable string network. 

Excitingly, the newly released PTA data~\cite{NANOGrav:2023hvm} can be explained by GWs originating from metastable cosmic string networks. The data show a preference for string tension values in the range of $G\mu_{cs}\sim 10^{-8}-10^{-5}$ (where the Newton's gravitational constant, $G= 6.7 \times 10^{-39}~\text{GeV}^{-2}$; hence, $G\mu_{cs}$ is a dimensionless quantity) for $\sqrt{\alpha}\sim 7.7-8.3$, with strong correlations between these two quantities~\cite{NANOGrav:2023hvm}. Importantly, these results are in full agreement with constraints obtained from Cosmic Microwave Background observations. Conversely, stable cosmic strings are not favored by the recent PTA results.

From the data, it is obtained that the $68\%$ credible region in the $G\mu_{cs}-\sqrt{\alpha}$ parameter plane overlaps with the third advanced LIGO–Virgo–KAGRA (LVK) bound~\cite{NANOGrav:2023hvm}. However, most of the $95\%$ credible region in the same parameter plane remains fully consistent with the data, favoring $G\mu_{cs}\lesssim 10^{-7}$ and $\sqrt{\alpha}\sim 8$~\cite{NANOGrav:2023hvm} (for example, from Eq.~\eqref{eq:alpha}, with $g=0.7$, a ratio, $M^\prime/M= 1.117$, of the two scales corresponds to $\sqrt{\alpha}\sim 8$). An interesting point to note is that $G\mu_{cs}\sim 10^{-7}$ approximately corresponds to $M\sim 10^{15}$ GeV, which aligns perfectly with the type-I seesaw contribution to neutrino mass and also matches the correct scale for inflation.

\color{black}
\section{Metastable cosmic strings \\from Pati-Salam Models}\label{sec:3}
First, we point out that within the Pati-Salam model, only three symmetry breaking chains can give rise to a metastable cosmic string network. To demonstrate this, first, we denote the various gauge groups as follows:
\begin{align*}
&G_{422}\equiv SU(4)_{C}\times SU(2)_{L} \times SU(2)_{R},
\\
&
G_{421}\equiv SU(4)_{C}\times SU(2)_{L} \times U(1)_R,
\\
&G_{3221}\equiv SU(3)_{C}\times SU(2)_{L} \times SU(2)_{R}\times U(1)_{B-L},
\\
&G_{3211}\equiv SU(3)_{C}\times SU(2)_{L} \times U(1)_{R}  \times U(1)_{B-L},
\\
&G_\mathrm{SM}\equiv SU(3)_{C}\times SU(2)_{L} \times U(1)_{Y},
\\
&G_\mathrm{31}\equiv SU(3)_{C}\times U(1)_\mathrm{em}.
\end{align*}

The breaking chains compatible with providing metastable cosmic strings are, therefore,
\begin{align}
\mathrm{Chain-I}:\; &G_{422}\xrightarrow[\langle\Phi_1\rangle]{m_\mathrm{red}} G_{3221}\xrightarrow[\langle\Phi_2\rangle]{m_\mathrm{blue}} G_{3211}\xrightarrow[\langle\Phi_{4,5}\rangle]{cs} G_\mathrm{SM}. \label{eq:chain1}
\\
\mathrm{Chain-II}:\;&G_{422}\xrightarrow[\langle\Phi_2\rangle]{m_\mathrm{blue}} G_{421}\xrightarrow[\langle\Phi_1\rangle]{m_\mathrm{red}} G_{3211}\xrightarrow[\langle\Phi_{4,5}\rangle]{cs} G_\mathrm{SM}. \label{eq:chain2}
\\
\mathrm{Chain-III}:\; &G_{422}\xrightarrow[\langle\Phi_{1,2}\rangle / \langle\Phi_3\rangle]{m_\mathrm{red},m_\mathrm{blue}} G_{3211}\xrightarrow[\langle\Phi_{4,5}\rangle]{cs}   G_\mathrm{SM}. \label{eq:chain3}
\end{align}
For each of these cases, the additional symmetry breaking stage, namely $G_\mathrm{SM}\to G_{31}$ is not shown explicitly. The representations (under the Pati-Salam group) of these fields playing role in symmetry breaking are $\Phi_1=(15,1,1)$, $\Phi_2=(1,1,3)$, $\Phi_3=(15,1,3)$, and $\Phi_4=(4,1,2)$ ($\Phi_5=(\overline 4,1,\overline 2)$). Note that $G_{422}\to G_{3211}$ can be obtained via the vevs of  two fields $\Phi_1(15,1,1)+\Phi_2(1,1,3)$ or by the vev of a single field $\Phi_3(15,1,3)$. 

The topological defects arising in these breaking chains are denoted by $m_\mathrm{red}$ (red monopole), $m_\mathrm{blue}$ (blue monopole), and $cs$ (cosmic string). The breaking of the gauge group $SU(4)_C\to SU(3)_C\times U(1)_{B-L}$ ($SU(2)_R\to U(1)_R$) leads to monopoles that carry both color and $B-L$ (only $R$) magnetic charges (charge) (which is referred to as the red (blue) monopole in Ref.~\cite{Lazarides:2019xai}). Finally, the last symmetry breaking scale before the electroweak breaking, i.e., $U(1)_R\times U(1)_{B-L}\to U(1)_Y$ leads to the formation of cosmic strings. Following the discussion above, through the quantum tunneling of the monopole-antimonopole pairs, strings eventually disappear if the monopole and cosmic string formation scales almost coincide. Our primary focus of this work is on these metastable cosmic string networks that may have formed in the very early universe, which emit gravitational waves that may have been observed in the PTAs.  

Since inflation must eliminate the unwanted monopoles, within the supersymmetric framework, there are different possibilities for achieving this successfully. The simplest scenario is the standard hybrid inflation. Since, in this case, the waterfall happens at the end of the inflation, the only consistent option is that inflation takes place at the $G_{3211}\to G_\mathrm{SM}$ breaking stage. Consequently, the waterfall leading to cosmic string formation is not inflated away, however, monopoles are. Another possibility could be implementing shifted hybrid inflation~\cite{Jeannerot:2000sv,Lazarides:2020zof}, where inflation can occur at an earlier stage. For example, shifted hybrid inflation in which symmetry breaking
$G_{421}\to G_{3211}$ proceeds along an inflationary trajectory can inflate away heavy monopoles (in principle, the same mechanism can be applied for $G_{3221}\to G_{3211}$ or $G_{422}\to G_{3211}$ scenario). 
    
\color{black}

\section{Model}\label{sec:4}
Although each of the scenarios discussed in the previous section is worth exploring, we focus on a concrete model in this work, as detailed in the following. As mentioned in the introduction, we work in the supersymmetric framework. 

In the Pati-Salam model with gauge symmetry $G_{422}$, the SM fermions belong to the following representation: 
\begin{align}
F_i&=(4,2,1)_i=Q_i(3,2,1/6)+L_i(1,2,-1/2),\\
\overline F_i&=(\overline 4,1,\overline 2)_i=u^c_i(\overline 3,1,-2/3)+d^c_i(\overline 3,1,1/3)+e^c_i(1,1,1)\nonumber\\
&+\nu^c_i(1,1,0).
\end{align}
Note that the fermionic multiplet, $\overline F_i$, additionally contains the right-handed neutrinos (RHNs), $\nu^c_i$, hence SM neutrinos naturally get tiny masses through type-I seesaw mechanism~\cite{Minkowski:1977sc, Gell-Mann:1979vob, Glashow:1979nm, Mohapatra:1979ia, Yanagida:1979as}. This same seesaw scale determines the cosmic string network formation scale, and as aforementioned, the new PTA data prefers this scale to be of order$\sim 10^{15}$ GeV.

The model we explore consists of the symmetry breaking chain given in Eq.~\eqref{eq:chain1}, i.e.,
\begin{align}
G_{422}\xrightarrow[\langle \Sigma\rangle]{M_X} G_{3221}\xrightarrow[\langle \Delta\rangle]{M^\prime} G_{3211}\xrightarrow[\langle H\rangle+ \langle \overline H\rangle]{M} G_\mathrm{SM} 
\xrightarrow[\langle h\rangle]{M_\mathrm{EW}} G_{31}.
\end{align}
Here, $M_X$ refers to the Pati-Salam breaking scale. 
From the discussion of the previous section, one finds that $M^\prime$ and $M$ are the monopole creation scale and cosmic string formation scale, respectively. At $M$, since $B-L$ symmetry is broken by two units, RHNs acquire their superheavy masses. Moreover, inflation, for which we employ the standard hybrid inflation, is also associated with this latter symmetry breaking scale. Formation of metastable cosmic string network showing consistency of recent PTA data requires $M^\prime \sim M\sim 10^{15}$ GeV. As for the Pati-Salam breaking scale, we choose, $10^{16}\;\mathrm{GeV}\lesssim M_X \lesssim10^{18}\;\mathrm{GeV}$.

The above symmetry breaking chain proceeds through the following set of Higgs representations:
\begin{align}
h&=(1,2,\overline 2)_H  =H_{u}(1,2,\frac{1}{2})+H_{d}(1,2,-\frac{1}{2}),
\\
\overline H&=(\overline 4,1,\overline 2)_H=u^{c}_{H}(\overline{3},1,-\frac{2}{3})+d^{c}_{H}(\overline{3},1,\frac{1}{3})\nonumber\\
&+e^{c}_{H}(1,1,1)+\nu^{c}_{H}(1,1,0),
\\
H&=(4,1,2)_H=\overline{u}^{c}_{H}(3,1,\frac{2}{3})+\overline{d}^{c}_{H}(3,1,-\frac{1}{3})\nonumber\\
&+\overline{e}^{c}_{H}(1,1,-1)+\overline{\nu}^{c}_{H}(1,1,0),
\\
\Delta&=(1,1,3)_H=\Delta_{-1}(1,1,-1)+\Delta_0(1,1,0)+\Delta_{1}(1,1,1),
\\
\Sigma&=(15,1,1)_H=\Sigma_0(1,1,0)+\Sigma_{2/3}(3,1,2/3)\nonumber\\
&+\Sigma_{-2/3}(\overline 3,1,-2/3)+\Sigma_8(8,1,0),
\end{align}
where decomposition of these multiplets under the SM gauge group are presented.

Within the SUSY context, a flat direction to obtain inflation naturally takes place in the R-symmetric scenario.  A gauge singlet superfield, $S=(1,1,1)$, plays the role of the inflation (the scalar component of it), which carries a full $U(1)_R$-charge (whereas, $H, \overline H, \Sigma, \Delta$ carry zero R-charge). Hence, the last intermediate scale symmetry breaking as well as inflation take place via the following superpotential: 
\begin{align}
W_\mathrm{Inflation}\supset \kappa S\left( H\overline H - M^2 \right).    
\end{align}
And the first two stages of symmetry breaking proceed through terms,\begin{align}
W_\mathrm{PS-breaking}\supset  \kappa^\prime S^\prime \left( \Sigma^2 - m^2_\Sigma \right) 
+ \kappa^{\prime\prime} S^{\prime\prime} \left( \Delta^2 - m^2_\Delta \right),
\end{align}
where $S^\prime, S^{\prime\prime}$ are also gauge singlets and carry full R-charge. In principle, terms involving $S, S^\prime, S^{\prime\prime}$ (each carrying full R-charge) all can mix and allow additional terms, which, for simplicity, are not considered in this work.

In the first breaking, the (true) Goldstones are $(3,1,2/3)+(\overline 3,1,-2/3)$; in the second breaking, Goldstones are $(1,1,1)+(1,1,-1)$; in the third breaking, the Goldstones are $(1,1,0)+(1,1,0)$. Therefore, due to R-symmetry,  the would-be Goldstones will be:
\begin{align}
&(8,1,0)\subset \Sigma,
\\
&(\overline 3,1,1/3)+(3,1,-1/3)\subset H,\overline H,
\\
&(\overline 3,1,-2/3)+(3,1,2/3)\subset H,\overline H, \Sigma,
\\
&(1,1,-1)+(1,1,1)\subset H,\overline H, \Sigma, \Delta \;.
\end{align}

To give masses to a set of submultiplets, we introduce
\begin{align}
&D_6=(6,1,1)_H= D_{3}(3,1,-\frac{1}{3})+\overline{D}_{3}(\overline{3},1,\frac{1}{3}),
\end{align}
which carries full R-charge (2 units) and allow the following interactions,
\begin{align}
W_\mathrm{mix}\supset HHD_6+\overline H \overline H D_6\;.      
\end{align}

Due to the above interactions, they give masses to $(3,1,-1/3) +(\overline 3,1,1/3)\subset H,\overline H$ of order $\langle H\rangle=\langle \overline H\rangle\sim 10^{15}$ GeV. A similar mechanism for the rest of the would-be Goldstones cannot be implemented.

Therefore, the rest of the would be Goldstones will acquire only SUSY breaking masses (due to $\langle S\rangle\sim M_\mathrm{SUSY}$). We also point out that extra Goldstones may result if no mixing terms is there between two multiplets carrying submultiplets with the same quantum numbers, namely $(\overline 3,1,-2/3)+(3,1,2/3)$  and $(1,1,-1)+(1,1,1)$. Teh following mixing terms can be written down that can also provide SUSY scale masses to these would-be Goldstones: 
\begin{align}
W_\mathrm{mix}\supset  H \overline H (\Sigma+\Delta) \frac{S}{\Lambda}\;.     
\end{align}

In summary, in the model under consideration, supermultiplets that reside at the SUSY scale are:
\begin{align}
&  
2\times \bigg\{ (3,1,2/3)+(\overline 3,1,-2/3) \bigg\}\nonumber\\
&
+2\times \bigg\{ (1,1,1)+(1,1,-1) \bigg\}
+1\times \bigg\{ (8,1,0) \bigg\}. 
\end{align}
The presence of these additional light states spoils the successful gauge coupling unification of the MSSM. Since gauge couplings are not necessarily unify in the Pati-Salam setup, one still has to worry about the perturbativity of the couplings at higher scales, which we discuss below.

Since the intermediate scales $M^\prime$ and $M$ are expected to be very high ($M^\prime\sim M\sim 10^{15}$ GeV) and can be somewhat close to the $M_X$ scale. Therefore, for the running of the gauge couplings,  it is a good approximation to consider 
a single step symmetry breaking at the high scale,
\begin{align}
G_{422}\xrightarrow{M_X} \mathrm{MSSM}\xrightarrow{M_\mathrm{SUSY}} \mathrm{SM}\;.    
\end{align}
The well-known $\beta$-coefficients~\cite{Jones:1981we,Machacek:1983tz} for the RGEs are~\cite{Arason:1991ic} $(a_1,a_2,a_3)_\mathrm{SM}=(41/10, -19/6, -7)$, and $(a_1,a_2,a_3)_\mathrm{MSSM}=(33/5, 1, -3)$. Moreover, for our scenario with the aforementioned additional states, we get $(a_1,a_2,a_3)=(61/5, 1, 2)_\mathrm{MSSM+extra}$. We run the corresponding RGEs from the low scale to the $M_X$ scale. By considering one-loop RGE analysis, we find that at $M_X=10^{16}$ GeV, the gauge couplings take the values $(g_1,g_2,g_3)=(4.08, 0.68, 1.97)$ for $M_\mathrm{SUSY}=3$ TeV and $(g_1,g_2,g_3)=(2.33, 0.67, 1.59)$ for $M_\mathrm{SUSY}=10$ TeV. On the other hand, if we set $M_X=10^{18}$ GeV, perturbativity of the couplings requires $M_\mathrm{SUSY}\geq 5000$ TeV. Therefore, in this setup, to ensure perturbativity it is safer to consider $M_\mathrm{SUSY}\geq 5000$ TeV (obtained from the crude estimation mentioned above).

Before closing this section, we write down the Yukawa part of the Lagrangian, which takes the following form, 
\begin{align}
W_Y\supset y F\overline F h + \frac{y_N}{\Lambda} \overline F\overline F H H, \label{eq:Majorana}
\end{align}
where $\Lambda$ denotes a cut-off scale such that $\Lambda > M_X$. 
The last term in the above superpotential provides Majorana mass for the RHNs when the last stage of the symmetry takes place.

\section{Details of Inflation and Baryon Asymmetry}\label{sec:5}
Minimal supersymmetric $\mu$-hybrid inflation employs a canonical K\"ahler potential and a unique renormalizable superpotential $W$ which respects a $U(1)\ R$ symmetry  as \cite{Dvali:1997uq},
\begin{equation}\label{SP}
W=\kappa S\left( H\overline H - M^2 \right)+\lambda S h^2,
\end{equation}
where $\kappa$ and $\lambda$ are dimensionless real parameters. The scalar part of the gauge singlet chiral superfield $S$ serves as the inflaton. The parameter $M$, which has mass dimensions, represents the non-zero vacuum expectation value (vev) of chiral superfields $H$ and $\overline{H}$.

The superpotential $W$ and superfield $S$ possess two units of $R$ charges, while the remaining superfields are assigned zero $R$ charges. Consequently, in the supersymmetric limit, the vev of the scalar component of superfield $S$ is zero. However, due to gravity-mediated supersymmetry breaking, the scalar component of $S$ acquires a non-zero vev proportional to $m_{3/2}$ as pointed out in~\cite{Dvali:1997uq}.

The last term in the superpotential, $\lambda S h^2$, effectively accounts for the $\mu$ term, where $\mu\sim(\lambda/\kappa) m_{3/2}$. This solution to the MSSM $\mu$ problem is described in \cite{Dvali:1997uq}.
The minimal canonical K\"ahler potential is given by 
 \begin{equation}\label{kahler}
 K_c=  |S|^2+|H|^2+|\overline{H}|^2 +|h|^2.
 \end{equation}
 Considering the effects of well-established radiative corrections \cite{Coleman:1973jx}, supergravity (SUGRA) corrections \cite{Linde:1997sj}, and the soft supersymmetry breaking terms \cite{Rehman:2009nq}, the inflationary potential, which arises along the D-flat direction with $|H|=|\overline{H}|=0$ and $h=0$, can be approximately expressed as follows:
 \begin{align}\label{potential}
V  &\simeq
\kappa^2 M^4 \left[ 1 + \left( \frac{M}{m_{P}}\right) ^{4}\frac{x^{4}}{2}+\frac{\kappa ^{2}}{8\pi ^{2}}F(x) + \frac{\lambda ^{2}}{4\pi ^{2}}F(y) \right. \notag \\
&\left.+ a\left(\frac{m_{3/2}\,x}{\kappa\,M}\right) + \left( \frac{M_S\,x}{\kappa\,M}\right)^2 \right],
\end{align}
where, $x=|S|/M$, $y=\sqrt{\gamma}\ x$, $M_S$ is the soft mass of the singlet. The parameter $\gamma$ is defined as $\gamma=\lambda/\kappa$, and $m_P$ stands for the reduced Planck mass.
The radiative corrections are described by the function,
\begin{align}
F(x)&=\frac{1}{4}\left[ \left( x^{4}+1\right) \ln \frac{\left( x^{4}-1\right)}{x^{4}}+2x^{2}\ln \frac{x^{2}+1}{x^{2}-1}\right. \notag\\
&\left.+2\ln \frac{\kappa ^{2}M^{2}x^{2}}{Q^{2}}-3\right],
\end{align}
and the coefficient of the soft SUSY-breaking linear term is defined as,
\begin{equation}
	a=2\vert A-2 \vert \cos \left(\arg S + \arg \vert A-2 \vert\right)\,.
\end{equation}
Both the linear term ($a$) and the mass-squared ($M^2_S$) soft SUSY-breaking terms in Eq.~\eqref{potential} are obtained in a gravity-mediated SUSY-breaking scheme. It's important to note that we will focus only on the real component of $S$, denoted as $\sigma=|S|/\sqrt 2$, where both the superfield and its scalar component are denoted by $S$\footnote{The imaginary component of $S$ is neglected here, which has been studied in \cite{Buchmuller:2014epa}.}.

At the end of the inflation epoch, the vacuum energy is converted into the energies of coherent oscillations of the inflaton S and the scalar field $\theta =\left(\delta H + \delta \overline{H} \right)/\sqrt{2}$,  which subsequently decay, giving rise to radiation in the universe. The $\mu$-term coupling $\lambda Sh^2$ in Eq.~\eqref{SP} leads to the inflaton's decay mostly into Higgsinos, with a decay width given by \cite{Lazarides:1998qx}
\begin{equation}
\Gamma_1=\frac{\lambda ^2 }{8 \pi }m_{inf},
\end{equation}
where $m_{inf}=\sqrt{2\kappa^2 M^2+M_S^2}$  represents the inflaton mass. The alternative decay channel for the inflaton is the decay to the right-handed neutrino through a dimension-5 operator $\beta_{ij}H H N_i^c N_j^c/M^*$,  which is another potential process. 
Heavy Majorana masses for the right-handed neutrinos are provided by the following  term 
\begin{equation}
M_{\nu^c_{ij}}=\beta_{ij}^{\prime}\frac{\langle H\rangle \langle H\rangle }{M^{*}}~\cdot 
\end{equation}
Also, Dirac neutrino masses of the order of the electroweak scale are obtained from the tree-level superpotential term  ${y_{\nu}}_{ij}\,N_{i}^c\,L_{j}\,H_{u}\to {m_{\nu_D}}_{ij}N N^c $ .  Thus, the neutrino sector is
\begin{eqnarray}
 W\supset  {m_{\nu_D}}_{ij}N_iN_j^c+ M_{\nu^c_{ij}}N_i^cN_j^c.   
\end{eqnarray}
 The small neutrino masses supported by neutrino oscillation experiments, are obtained by integrating out the heavy right-handed neutrinos and read as
\begin{equation}
m_{\nu_{D_{\alpha\beta}}}=-\sum_{i}y_{\nu_{i\alpha}} y_{\nu_{i\beta}}\frac{v_{u}^2}{M_i}. \label{mneu1}
\end{equation}
The neutrino mass matrix $m_{\nu_{D_{\alpha\beta}}}$ can be diagonalized by a unitary matrix $U_{\alpha i}$ as $m_{\nu_{D_{\alpha\beta}}} =  U_{\alpha i} U_{\beta i} \hat{m}_{\nu_{D_{i}}}$, where $\hat{m}_{\nu_{D_{i}}}$ is a diagonal
mass matrix $m_{\nu_D} = {\rm diag}(m_{\nu_{1}}, m_{\nu_{2}}, m_{\nu_{3}})$ and $M_{i}$ represent the eigenvalue of mass matrix  $M_{\nu^c_{ij}}$. Then the decay width for the inflaton decay into RH neutrinos is given by 
    \begin{equation}
\Gamma_2=\frac{m_{inf}}{8 \pi }\left(\frac{M_N}{M}\right)^2\left(1-\frac{4M_N^2}{m_{inf}^2}\right)^{1/2},
\end{equation}
The reheat temperature $T_R$ is estimated to be \cite{Kolb:1990vq}:

\begin{equation}\label{reheat}
T_R\approx\sqrt[4]{\frac{90}{\pi^2{g_*}}} \sqrt{{\Gamma } {m_P}} , \quad \Gamma=\Gamma_1+\Gamma_2
\end{equation}
where $g_*$ takes the value 228.75 for MSSM. The lepton asymmetry is generated through right-handed neutrino decays. The lepton number density
to the entropy density  in the limit $T_R < M_{1}\equiv M_{N}\leq m_{\text{inf}} /2 \leq M_{2,3}$ is defined as
\begin{equation}
\frac{n_{L}}{s}\sim \frac{3}{2} \left(\frac{\Gamma_2}{\Gamma}\right)\frac{T_{R}}{m_{\text{inf}}}\epsilon_{cp}~,
\end{equation}
where $\epsilon_{cp}$ is the CP asymmetry factor and is generated from the out of equilibrium decay of lightest right-handed neutrino and is given by,
\begin{equation}
\epsilon_{cp}=-\frac{3}{8\pi}\frac{1}{\left({y_{\nu}}{y_{\nu}}^{\dagger}\right)_{11}}\sum_{i=2,3}\operatorname{Im} \left[\left({y_{\nu}}{y_{\nu}}^\dagger\right)_{1i}\right]^2\frac{M_{N}}{M_i},
\end{equation}
Assuming a normal hierarchical pattern of light neutrino masses, the CP asymmetry factor, $\epsilon_{cp}$, becomes 
\begin{equation}
\epsilon_{cp} = \frac{3}{8\pi}\frac{M_{N} m_{\nu_{3}}}{v_{u}^2}\delta_{\rm eff}, 
\end{equation}
where $m_{\nu_3}$ is the mass of the heaviest light neutrino, $v_{u}=\langle H_u \rangle $ is the vev of the up-type electroweak Higgs and $\delta_{\rm eff}$ is the CP-violating phase. A successful baryogenesis is usually achieved through the sphaleron process where an
initial lepton asymmetry, $n_L/s$ is partially converted into the baryon asymmetry as $n_B/s \sim 0.35\,n_L/s$ \cite{Harvey:1990qw}. From the experimental value of the baryon to photon ratio $n_B \approx (6.1\pm 0.4) \times 10^{-10}$ \cite{ParticleDataGroup:2020ssz},  the
required lepton asymmetry is found to be
\begin{eqnarray}
\mid n_L/s\mid\approx\left(2.67-3.02\right)\times 10^{-10}.
\end{eqnarray}
In the numerical estimates discussed below we take $m_{\nu_3} = 0.05$ eV and $v_u = 174$ GeV, while assuming large $\tan \beta $. The non-thermal production of lepton asymmetry, $n_{L}/s$, is given by the following expression  
\begin{align}\label{bphr}
\frac{n_L}{s} \lesssim 3 \times 10^{-10} \left(\frac{\Gamma_2}{\Gamma}\right)\left(\frac{T_R}{m_{\text{inf}}}\right)\left(\frac{M_{N}}{10^6 \text{ GeV}}\right)\left(\frac{m_{\nu_3}}{0.05 \text{ eV}}\right),
\end{align}
with $M_{1} \gg T_R $.  To ensure inflationary predictions are in line with leptogenesis, we employ Eq. \eqref{bphr} for our numerical analysis.

\section{Numerical results} \label{CMB}

The prediction for the various inflationary parameters are calculated using the standard slow-roll parameters,
\begin{align}
\epsilon &= \frac{1}{4}\left( \frac{m_P}{M}\right)^2
\left( \frac{V'}{V}\right)^2, \,\,\,
\eta = \frac{1}{2}\left( \frac{m_P}{M}\right)^2
\left( \frac{V''}{V} \right),\nonumber\\ 
\xi^2 &= \frac{1}{4}\left( \frac{m_P}{M}\right)^4
\left( \frac{V' V'''}{V^2}\right).
\label{slowroll}
\end{align}
In the above, prime denotes the derivative with respect to $x$. Moreover, the scalar spectral index $n_s$, the tensor-to-scalar ratio $r$, and the running of the scalar spectral index $\alpha_{s}\equiv dn_s / d \ln k$, in the slow-roll approximation are given by,
\begin{align} \label{nsr}
n_s\simeq 1+2\,\eta-6\,\epsilon, \,\,\,
r \simeq 16\,\epsilon, \,\,\,
\alpha_{s} \simeq 16\,\epsilon\,\eta
-24\,\epsilon^2 - 2\,\xi^2,
\end{align}
with $n_s = 0.9665 \pm 0.0038$ \cite{Planck:2018jri} in the $\Lambda$CDM model.  The amplitude of the scalar power spectrum is given by,
\begin{align}
A_{s}(k_0) = \frac{1}{24\,\pi^2\,\epsilon(x_0)}
\left( \frac{V(x_0)}{m_P^4}\right),  \label{curv}
\end{align}
which at the pivot scale $k_0 = 0.05\, \rm{Mpc}^{-1}$ is given by $A_{s}(k_0) = 2.137 \times 10^{-9}$, as measured by Planck 2018 \cite{Planck:2018jri}.
And the number of e-folds, $N_0$, is given by,
\begin{align}\label{Ngen}
N_0 = 2\left( \frac{M}{m_P}\right) ^{2}\int_{x_e}^{x_{0}}\left( \frac{V}{%
	V'}\right) dx,
\end{align}
where $x_0 \equiv x(k_0)$ and
$x_e$ are the field values at the pivot scale $k_0$ and at the end of inflation, respectively. 
The value of $x_e$ is determined by the breakdown of the slow-roll approximation. Finally, the number of e-folds, $N_{0}$, can be written in terms of the reheat temperature~\cite{Liddle:2003as} (assuming a standard thermal history),
\begin{align}\label{efolds}
N_0=53+\dfrac{1}{3}\ln\left[\dfrac{T_R}{10^9 \text{ GeV}}\right]+\dfrac{2}{3}\ln\left[\dfrac{\sqrt{\kappa}\,M}{10^{15}\text{ GeV}}\right].
\end{align}

\begin{figure*}[t]
	\centering
	\subfloat[\label{sub:Ms_a}]
	{{\includegraphics[width=0.5\textwidth]{"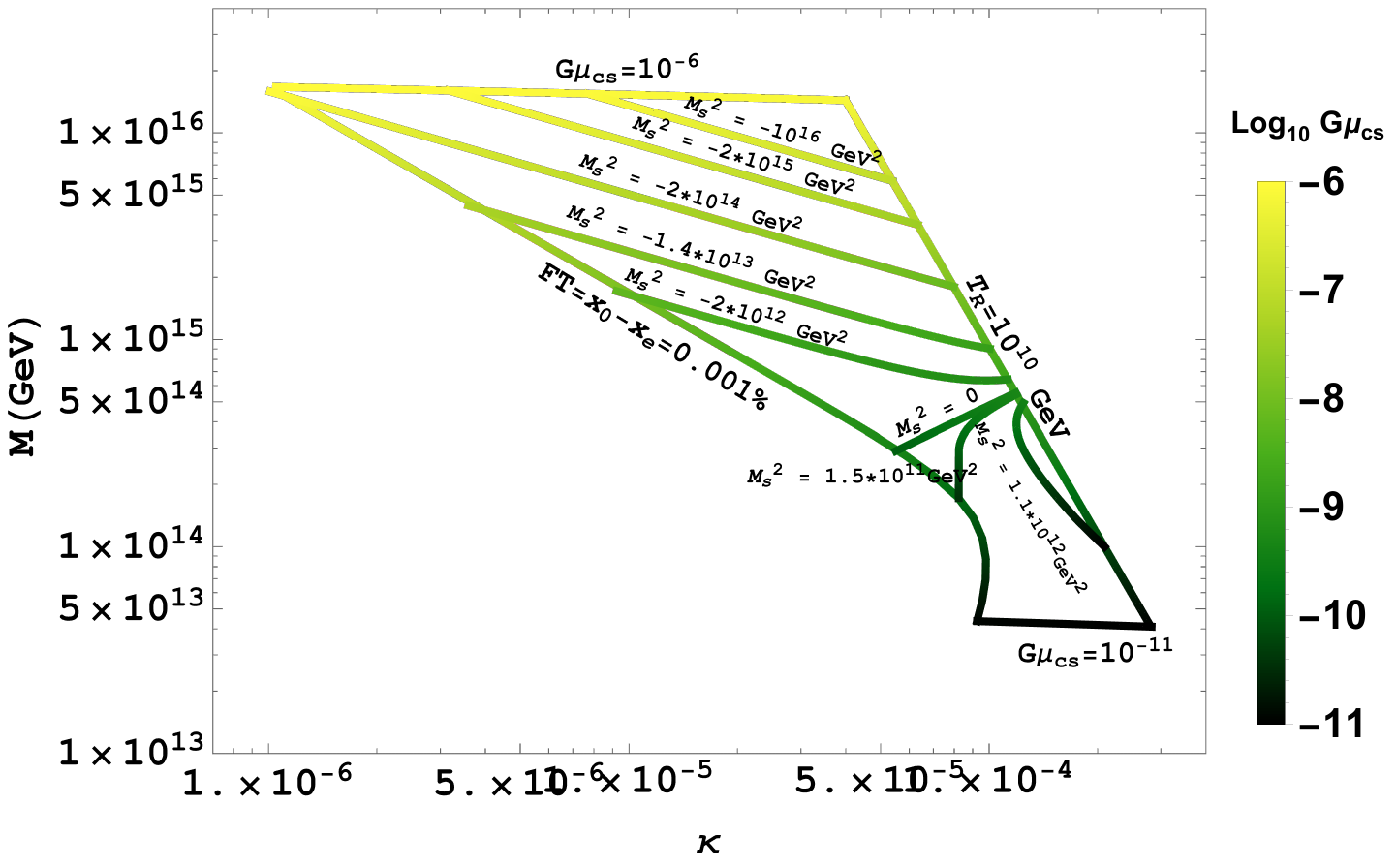"} }}%
	\subfloat[\label{sub:m32_a}]
	{{ 	\includegraphics[width=0.5\textwidth]{"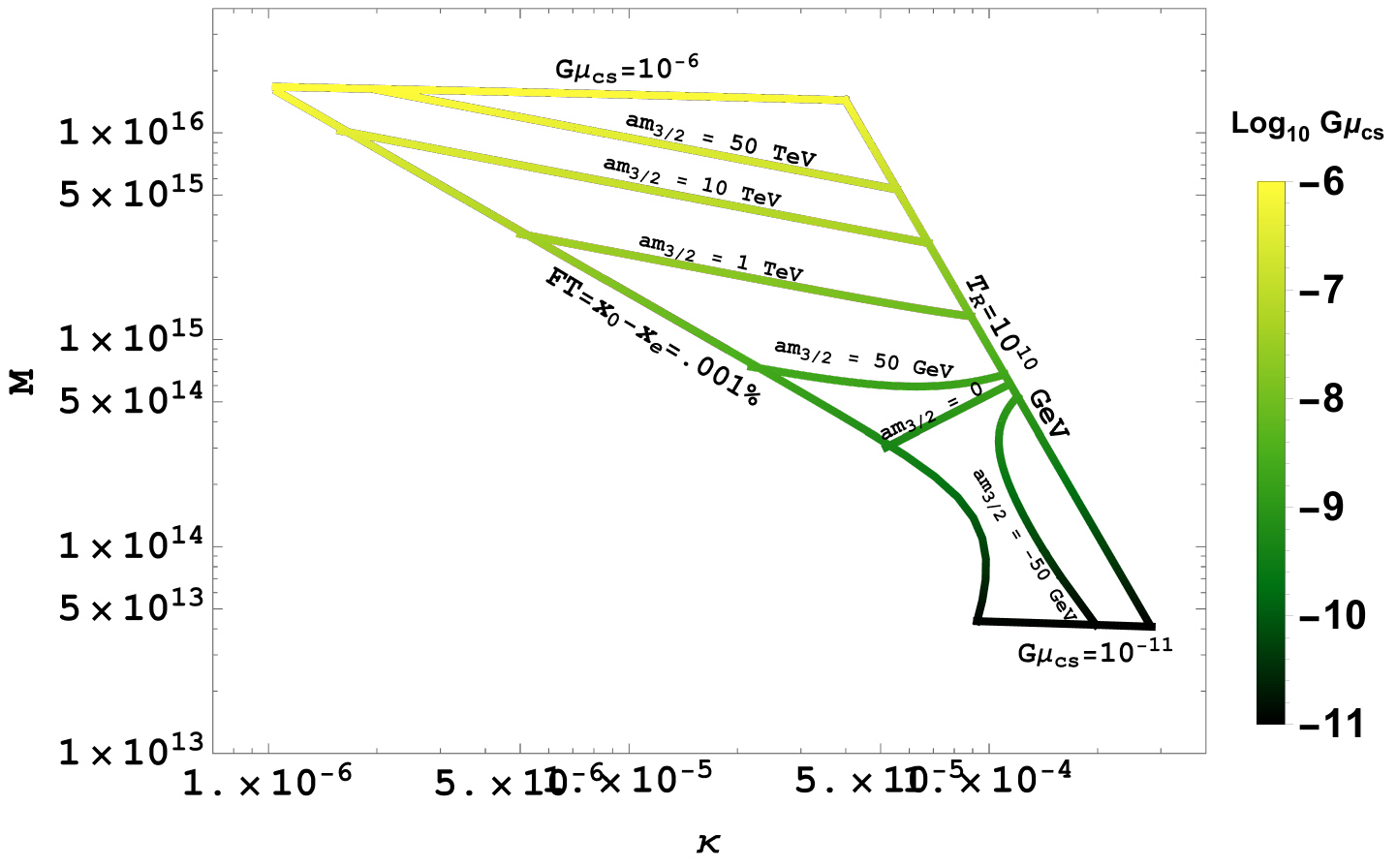"} }}%
	\caption{The relationship between the symmetry breaking scale ($M$) and the coupling strength ($\kappa$) is depicted. The reheat temperature ($T_R$) is bounded between a maximum of $10^{10}$ GeV and a minimum of $2.3\times10^6$ GeV and fine tuning bound of $0.001\%$ is applied. An Avocado colored vertical bar represents the variation of the dimensionless string tension $G\mu_{cs}$  from $10^{-6}$ to $10^{-11}$. The inside mesh shows the variation of soft mass $M_S^2$ in \ref{sub:Ms_a} (left panel) and that of  $m_{3/2}$ in \ref{sub:m32_a} (right panel).}
	\label{fig:gravitino}
\end{figure*}

\begin{figure*}[t]
	\centering
	\subfloat[\label{sub:tensor}]
	{{\includegraphics[width=0.5\textwidth]{"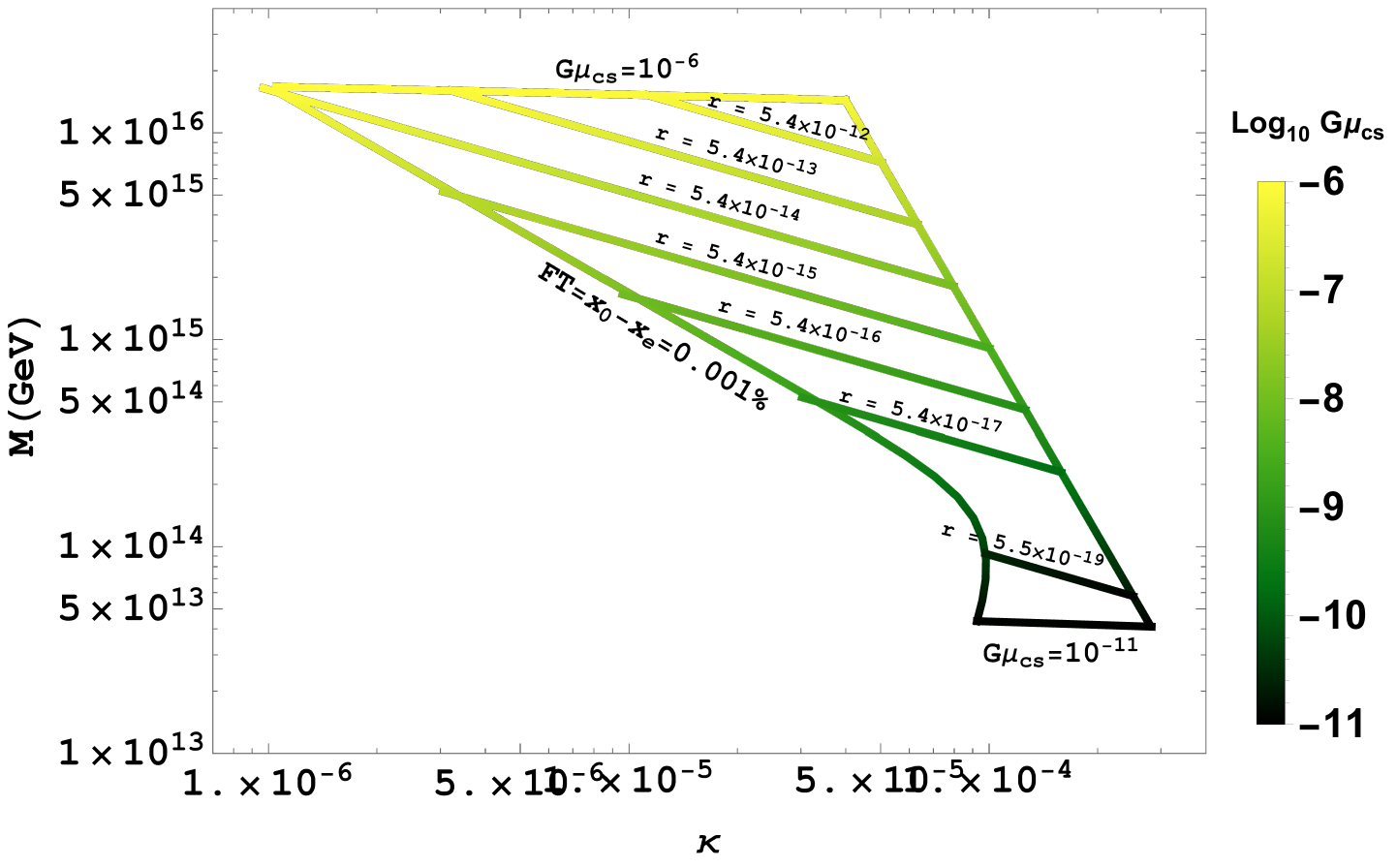"} }}%
	\subfloat[\label{sub:Reheat}]
	{{ 	\includegraphics[width=0.5\textwidth]{"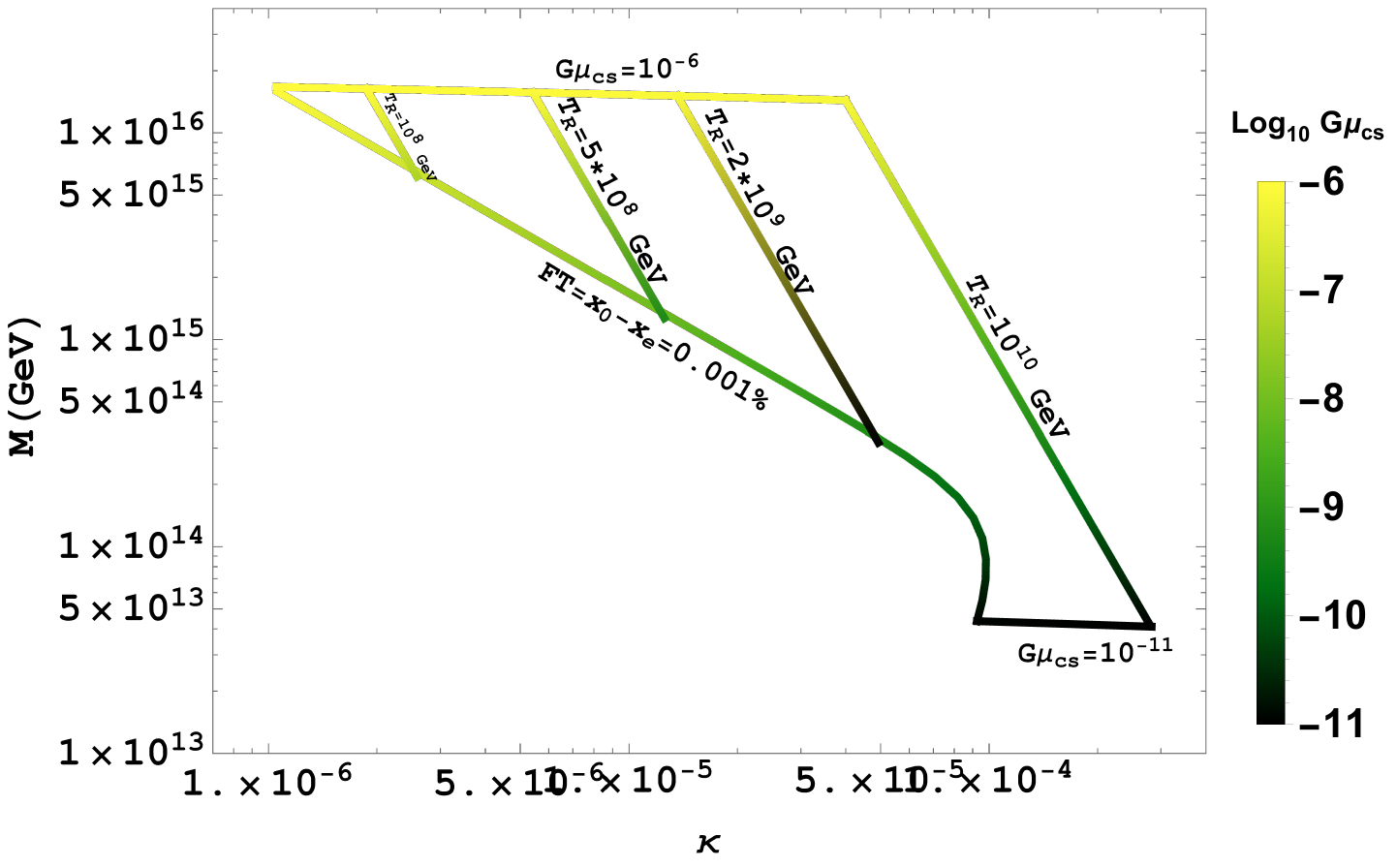"} }}%
	\caption{The relationship between the symmetry breaking scale ($M$) and the coupling strength ($\kappa$) is depicted. The reheating temperature ($T_R$) is bounded between a maximum of $10^{10}$ GeV and a minimum of $2.3\times10^6$ GeV and fine tuning bound of $0.001\%$ is applied. An Avocado colored vertical bar represents the variation of the dimensionless string tension $G\mu_{cs}$  from $10^{-6}$ to $10^{-11}$. The inside mesh shows the variation of tensor-to-scalar ratio $r$ in left panel Fig \ref{sub:tensor}, and the right panel the reheating temperature $T_R$ in Fig \ref{sub:Reheat}.}
	\label{fig:tensor_Tr}
\end{figure*}

In our numerical analysis we have seven independent key parameters: $\kappa$, $M$, $am_{3/2}$, $M_S^2$, $x_0$, $x_e$, and $M_N$. These parameters are subject to five essential constraints:
\begin{itemize}
    \item The amplitude of the scalar power spectrum, denoted as $A_s(k_0)$, with a specific value of $2.137\times 10^{-9}$ (as given in Eq~\eqref{curv})
    \item   The scalar spectral index, represented by $n_s$, which holds a fixed value of $0.9665$ \cite{Planck:2018jri}.
    \item  The end of inflation, determined by the waterfall mechanism, with the condition that $x_e=1$.
    \item The number of e-folds, denoted as $N_0$ in Eq.~\eqref{Ngen} is defined in terms of $T_R$ by Eq.~\eqref{efolds}. 
    \item The observed value of the baryon asymmetry, which translate a boud on lepton asymmetry expressed as $n_L/s$, which takes the specific value of $3\times 10^{-10}$ (as given in Eq.~\eqref{bphr}).
\end{itemize}

These constraints are really important and play a crucial role for figuring out different possibilities in the model's predictions. When we take these constraints into account, we end up with two independent parameters that we can vary freely. We choose these parameters to be $am_{3/2}$ and $M_S^2$. By fixing one of these parameters, we can then explore the variations of the other.

We displayed our numerical calculations in Fig. \ref{fig:gravitino}, which illustrates how the parameters change across the $\kappa-M$ plane. During our analysis, we kept the scalar spectral index at the central value allowed by Planck, which is $n_{s}=0.9655$ \cite{Planck:2018jri}. To make sure the SUGRA expansion doesn't go out of control, we required  $S_{0}\leq m_{p}$ in our parametric space.  Further, we restrict $M\leq 2\times10^{16}$ GeV and $T_{R}\leq 10^{10}$ GeV to avoid the gravitino problem \cite{Ellis:1984eq, Khlopov:1984pf}. Note that, although the recent PTA data prefers $M\sim 10^{15}$ GeV, in presenting our results, we try to be as general as possible and vary this scale in the range $10^{13}\mathrm{GeV}\lesssim M \lesssim 10^{16}\mathrm{GeV}$.  The explored parameter space yields a reheating temperature within the range of $(10^6-10^9)$ GeV. \footnote{In our model, the reheating temperature lies within the range of $(10^6-10^9)$ GeV. The constraints on the reheating temperature $T_R$ and gravitino mass $m_{3/2}$ arising from the Big Bang can be readily met within the parameter space discussed here, for both scenarios, namely, gravitino as stable and  unstable particle. For more details see \cite{Afzal:2022vjx}.}. We further restrict our numerical results by imposing the following conditions
\begin{equation}
	m_{inf}\geq2M_{N}, \qquad M_{N}\geq10 T_{R},
\end{equation}
which ensures successful reheating with non-thermal leptogenesis. The boundary curves in Fig.~\ref{fig:gravitino} represent; $M = 2 \times 10^{16}$ GeV, $T_R = 10^{10}$ GeV, $ FT = 0.001\%$ and $G \mu_{cs}= 10^{-11}$ constraints. In the left panel of Fig.~\ref{sub:Ms_a}, we explore the variation of $am_{3/2}$ across a range of $M_S^2$ values, spanning from $2\times10^{12}\, \text{GeV}^2$ to $-2\times10^{17}\, \text{GeV}^2$. Similarly, the right panel of Fig.~\ref{sub:m32_a} demonstrates the variation of $M_S^2$ while keeping $m_{3/2}$ fixed within the interval of $0$ to $730\, \text{TeV}$ ($0$ to $155$~GeV) for the cases where $a$ is $1$ or $-1$.

In order to achieve a red-tilted scalar spectral index consistent with Planck-2018 data, at least one of the two parameters, $M_S^2$ or $am_{3/2}$, is expected to be negative \cite{Ahmed:2022vlc,Ahmed:2022thr}. The scalar spectral index, in the limit  $x_0 \sim 1$, can be approximated in the following way:
\begin{equation}\label{ns}
n_s \simeq 1+ \left(\dfrac{m_P}{M}\right)^2 \left( 2\left(\dfrac{M_S}{\kappa M}\right)^2 +3\dfrac{\kappa^2 }{8\pi^2}F''(x_0)\right).
\end{equation}
In the above expression,
when $M_S^2$ term is dominant,  we obtain,
\begin{equation}
  \left(\dfrac{M_S}{\kappa M}\right)^2 \simeq -\frac{(1-n_s)}{2} \left(\dfrac{M}{m_P}\right)^2.
\end{equation}
Hence, the consistent pattern of the curves in Fig. \ref{sub:Ms_a} across most of the upper region, displaying $M \propto \kappa^{-1/2}$ behavior, can be readily comprehended when considering constant values of $M_S^2$. Conversely, the $M \propto \kappa$ behavior observed near the curve where $M_S \sim 0$ can be attributed to the predominant radiative term within Eq. \eqref{ns}.

Regarding the behavior of the corresponding curves in Fig. \ref{sub:m32_a} with constant values of $am_{3/2}>0$, a competition arises among the soft SUSY breaking terms in $\epsilon(x_0)$ to meet the constraint imposed by $A_s$ in Eq. \eqref{curv}. This observation, combined with Eq. \eqref{ns}, leads to the emergence of a $M \propto \kappa^{-1/3}$ behavior. This behavior aligns with the curves exhibited in the upper region of Fig. \ref{sub:m32_a}."

In scenarios where $M_S^2>0$, as we depart from the $M_S \sim 0$ curve in Fig \ref{sub:Ms_a}, the radiative corrections compete  with the $M_S^2$ term in Eq.~\eqref{ns}. Consequently, we discern that the parameter $M$ varies in proportion to $\kappa^{-2}$, as clearly observed in the lower region of Fig \ref{sub:Ms_a}.

Now, focusing on the corresponding region displayed in Fig. \ref{sub:m32_a}, and holding constant values of $am_{3/2}>0$, in order to satisfy the constraint on the amplitude of the scalar power spectrum, $A_s$, as delineated in Eq. \eqref{curv}, the contributions from both soft supersymmetry (SUSY) breaking and radiative corrections become comparable within the expression for $\epsilon(x_0)$. This behavior characterized by $M\propto \kappa^{-3}$ for the curves featured in the lower part of Fig. \ref{sub:m32_a}.

In Figure \ref{fig:tensor_Tr}, in the $\kappa-M$ plane, the left panel depicts the variation of the tensor-to-scalar ratio $r$, while the right panel illustrates the variation of the reheating temperature $T_R$. The predicted range of the tensor-to-scalar ratio is tiny and lies in the range $r \sim 5\times10^{-11}-5\times10^{-21}$, see Fig.~\ref{sub:tensor}. The various curves with constant values of $r$ shows $M\propto \kappa^{-1/2}$ behavior, as can be deduced from Eq.~\eqref{curv}:
\begin{equation}
r \sim \frac{2}{3\,\pi^2 A_s(k_0)}\frac{\kappa^2 M^4}{m_P^4}.
\end{equation}
Similarly in Fig \ref{sub:Reheat}, the curves with fixed values of reheat temperature in the range ranging between $T_R\in [2 \times10^6-10^{10}]$~GeV follow $M \propto \kappa^{-3}$ behavior obtained from Eq.~\eqref{reheat}. All the solutions obtained in our analysis satisfy the leptogenesis constraint, as depicted in the Eq.~\eqref{bphr}.

\begin{figure}[th!]
\begin{center}
\includegraphics[width=0.5\textwidth]{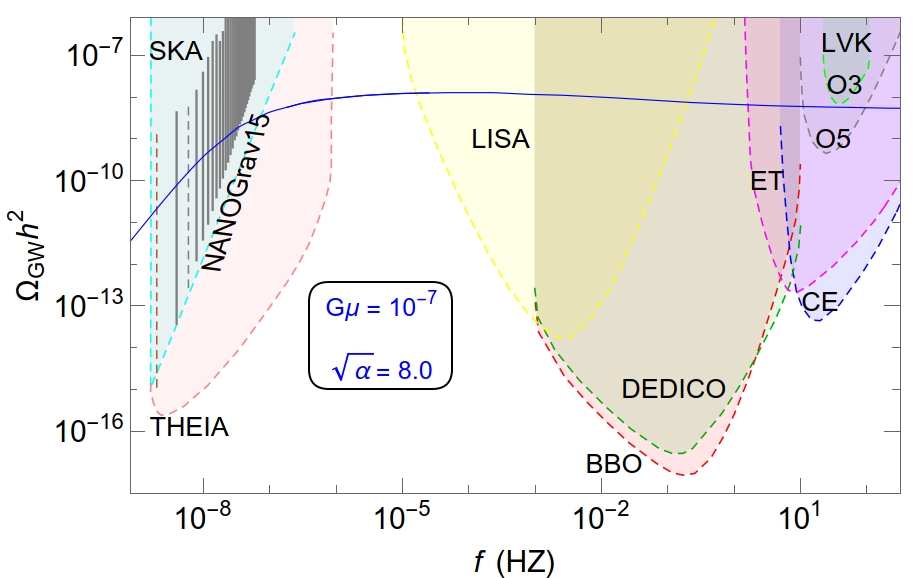}
\caption{
Gravitational wave signal from metastable cosmic strings explaining the recent PTA result~\cite{NANOGrav:2023hvm}. The current LVK~\cite{LIGOScientific:2021nrg,KAGRA:2021kbb} bound (O3) on cosmic string networks is depicted. Future sensitivities of various upcoming experiments, such as (LISA~\cite{LISA:2017pwj}, SKA~\cite{Janssen:2014dka}, THEIA~\cite{Theia:2017xtk},  ET~\cite{Sathyaprakash:2012jk} and CE~\cite{LIGOScientific:2016wof}, BBO~\cite{Corbin:2005ny}, and DECIGO~\cite{Seto:2001qf} are also illustrated. See text for details.
}\label{fig:GWNG15}
\end{center}
\end{figure}

Furthermore, as mentioned before, the gauge-breaking scale, denoted by $M$, is related to the cosmic string tension parameters $\mu_{cs}$. The vertical avocado bar in the plot indicates the range of the string tension parameter, $G\mu_{cs}$, which for making the plots, is varied in the range $\sim 10^{-6}$ to $10^{-11}$. As pointed out in Sec.~\ref{sec:2},  the third advanced LVK bound rules out regions of the parameter space for $G\mu_{cs}> 10^{-7}$~\cite{NANOGrav:2023hvm}, whereas the recent PTA data strongly suggests $G\mu_{cs}\sim 10^{-7}$ for $\sqrt{\alpha}$ close to 8~\cite{NANOGrav:2023hvm}. This LVK bound is slightly stronger than the constraints arising from CMB, which leads to $G\mu_{cs}\lesssim 1.3\times 10^{-7}$  \cite{Planck:2018vyg,Planck:2018jri}. Intriguingly, the entire region of the parameter space shown in all these plots above, with $G\mu_{cs}< 10^{-7}$ in a broad spectrum of frequencies, will be fully probed by several gravitational wave observatories. This is explicitly depicted in Fig.~\ref{fig:GWNG15} for an example parameter point with $(G\mu_{cs}, \alpha^{1/2})= (10^{-7}, 8)$, which explains the recent PTA data. For making this plot, we followed the procedure explained in Ref.~\cite{Buchmuller:2021mbb}.

\section{Conclusions}\label{sec:con}

In conclusion, we have investigated promising pathways for model building that connect the Pati-Salam model to the Standard Model, uncovering noteworthy cosmological implications along the way. The breaking pattern $G_{422}$ to $G_{3211}$ through $G_{3221}$ produce monopoles. To circumvent the issue of monopoles, we adopted the standard SUSY hybrid inflation at $G_{3211}$ scale, ensuring compatibility with matter-antimatter asymmetry obtained through leptogenesis. Our model not only produces a scalar tilt that aligns with the Planck 2018 constraints but also exhibits small tensor modes beyond the scope of upcoming CMB experiments. The breaking of $G_{3211}$ at the end of inflation leads to the cosmic strings, which eventually disappear due to the quantum tunneling of the monopole-antimonopole pairs, confirming the formation of metastable cosmic string network. The stochastic gravitational wave background produced by this network of metastable strings is compatible with the gravitational waves observed recently by pulsar timing array  experiments, including NANOGrav, CPTA, EPTA, InPTA, and PPTA. Furthermore, this gravitational wave spectrum remains within the detection capabilities of both existing and future ground-based and space-based experiments.

\subsection*{Acknowledgments}
T.A.C would like to thank the High Energy Theory Group in the Department of Physics and Astronomy at the University of Kansas for their hospitality and support. The work of S.N is supported by the United Arab Emirates University (UAEU) under UPAR Grant No. 12S093.  S.S\; would like to thank Qaisar Shafi for discussion and Kevin Hinze for his help in preparing figure 3.

\color{black}

\bibliographystyle{style}
\bibliography{references}
\end{document}